\documentclass[twoside]{article}
\usepackage{fleqn,espcrc2}

\usepackage{graphicx}
\usepackage[figuresright]{rotating}

\newcommand{\be}{\begin{equation}}
\newcommand{\ee}{\end{equation}}
\newcommand{\bea}{\begin{eqnarray}}
\newcommand{\eea}{\end{eqnarray}}
\newcommand{\ba}{\begin{eqnarray}}
\newcommand{\ea}{\end{eqnarray}}

\newcommand{\AmS}{{\protect\the\textfont2
  A\kern-.1667em\lower.5ex\hbox{M}\kern-.125emS}}

\title{Exact results and approximation schemes for the Schwinger Model with
  the Overlap Dirac Operator\thanks{The authors have been supported in part under 
                 DOE grant DE-FG02-91ER40676.}}
\author{L.~Giusti, C.~Hoelbling, C.~Rebbi\\
[0.2cm]
Boston University - Department of Physics, 590 
Commonwealth Avenue, Boston MA 02215.}
\begin{document}

\begin{abstract}
We propose new techniques to implement numerically the
overlap-Dirac operator which exploit the physical properties of the
underlying theory to avoid nested algorithms. We test these procedures 
in the two-dimensional Schwinger model and the results are 
very promising. We also present a detailed computation of the 
spectrum and chiral properties of the Schwinger Model 
in the overlap lattice formulation.
\vspace{1pc}
\end{abstract}

\maketitle
The overlap \cite{HN} formulation of lattice 
fermions provides a definition of a lattice Dirac
operator $D$ which avoids the doubling problem and preserves the
relevant symmetries of the continuum theory, most notably chiral
symmetry in the limit of vanishing fermion mass.  Unfortunately, this
welcome development has come at a price: the numerical calculation of
the matrix elements of the propagator $D^{-1}$ and the inclusion of
${\rm Det}(D)$ in the measure entail a substantially increased
computational burden, which severely constrains the maximum lattice
size for viable simulations. In this talk we present 
results of an exploratory study for a numerical approximation which exploits the
physical properties of the system. It proceeds through the projection over a
subspace which has a substantially reduced number of degrees of freedom
but still captures the relevant non-perturbative long range properties of the
model.  

The Schwinger model is an ideal laboratory to test our approximation:
it has many features in common with QCD, most notably chiral symmetry,
and it is simple enough that it is possible to compute $D^{-1}$ and 
${\rm Det}(D)$ exactly within the overlap formulation.
Our primary goal is to compare the results of our approximation 
to exact results. Nevertheless the results of the exact
calculations are interesting per se, because they are
more extensive than what, to the best of our knowledge, has been obtained
up to now and validate in an impressive manner the advantages of 
Neuberger's formulation of lattice fermions. This talk is based 
on a work which will be the subject of a forthcoming publication
\cite{NoiSch}, where all the notations used here and 
the details of the simulations are reported. 

\section{The Schwinger Model with Overlap Fermions}
The Euclidean action of the Schwinger Model in the 
overlap regularization is given by 
\ba
S_L & = & \beta \sum_{x,\mu<\nu}\left[ 1- \mbox{Re}\, U_{\mu\nu}(x)\right]
\label{eq:sg}   \\
& + & 
\sum_{i=1}^{N_f}\sum_{x,y} \bar{\psi}_i(x)\left[ \, 
(1-\frac{ma}{2}) D (x,y) + m \right]\, \psi_i(y)\nonumber
\ea   
where $U_{\mu\nu}(x)$ is the standard Wilson plaquette, 
$\psi_i$ are $N_f$ fermions fields with mass $m$ and  
$\beta = 1/(ag)^2$, $g$ being the bare coupling constant.
$D$ is the massless overlap-Dirac operator
\ba\label{eq:opneub}
D &=& \frac{1}{a}\left( 1 + X\frac{1}{\sqrt{X^\dagger X}}\right)\nonumber\\
X &=& D_W -\frac{1}{a}\; ,
\ea
where $D_W$ is the Wilson-Dirac operator and $a$ is the lattice spacing. 
The Neuberger-Dirac operator 
satisfies the $\gamma_5$-Hermiticity condition
\be\label{eq:gamma5H}
D^{\dagger} = \gamma_5 D \gamma_5
\ee
and the Ginsparg-Wilson (GW) relation \cite{GW} 
\be\label{eq:GW}
\gamma_5 D^{-1} + D^{-1} \gamma_5 = a \gamma_5 
\ee
The latter equation implies a continuous
symmetry of the fermion action in Eq.~(\ref{eq:sg}) at finite lattice spacing 
\cite{luscher}
\be\label{eq:luscher}
\delta \psi =  \hat \gamma_5 \psi \qquad \delta \bar \psi =  \bar \psi \gamma_5  
\ee
where $\hat \gamma_5 = \gamma_5(1-aD)$, which may be interpreted 
as a lattice form of the chiral symmetry.
The chiral transformations in Eq.~(\ref{eq:luscher})
lead to Ward identities analogous to the continuum 
ones and the anomaly term arises from the 
non-invariance of the fermion integral measure \cite{luscher}.
The corresponding flavor non-singlet chiral transformations are defined
including a flavor group generator in Eq.~(\ref{eq:luscher}). 
They imply the 
Axial Ward identities
\be\label{eq:AWI}
\partial_\mu A^c_\mu =   2 m P^c +O(a^2)
\ee
where the axial current and the pseudoscalar densities are
\ba
A^c_\mu = \bar\psi\frac{\lambda^c}{2}\gamma_\mu \gamma_5 \psi\; 
\qquad P^c = \bar\psi\frac{\lambda^c}{2}\gamma_5 \psi
\ea
with $\psi = (\psi_i,\dots\psi_{N_f})$,
and $\lambda^{c}$ are the generators of the 
$SU(N_f)$ flavor group. In two dimensions the Axial
current is related to the Vector one as 
\be
A^c_\mu(x) = -i \epsilon_{\mu\nu} V^c_\mu(x) 
\ee
where $\epsilon_{\mu\nu}$ is the totally antisymmetric tensor.
\begin{figure}[ht]
\includegraphics[height=6.5cm,width=7.5cm]{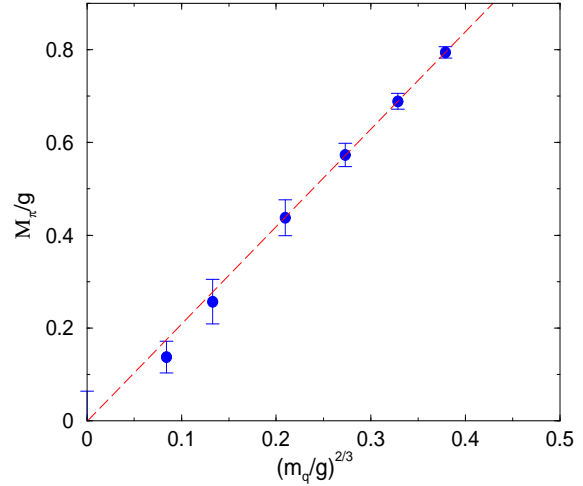}
\vspace{-40pt}
\caption{\label{fig:mpiNf2}
$M_\pi/g$ vs $(m/g)^{2/3}$ for the full operator for 
$N_f=2$.}
\end{figure} 
\begin{figure}[ht]
\includegraphics[height=6.5cm,width=7.5cm]{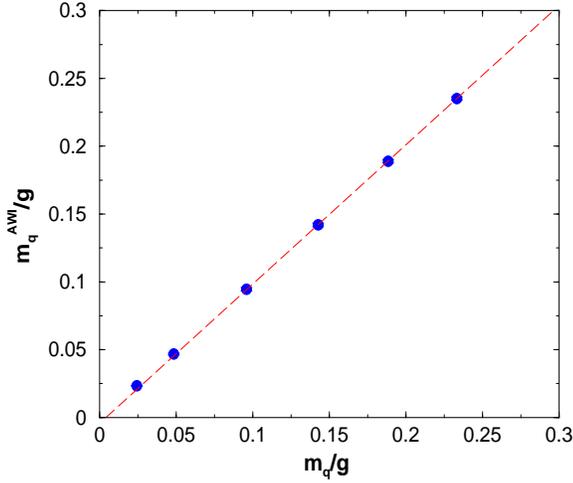}
\vspace{-40pt}
\caption{\label{fig:rhoNf2}
$m^{AWI}/g$ vs $m/g$ for the full operator for $N_f=2$.}
\end{figure} 

In the $N_f=2$ massless model, there is a triplet of massless
particles ($\pi$) and one massive singlet particle ($\eta$)
with mass $(M_\eta/g)^2=2/\pi$. The correlation functions
of the vector currents are the same as for free particles.
For small masses the corrections to the massless 
results can be obtained by ``chiral perturbation'' theory 
\cite{ghl} which to first order gives
\ba
\label{eq:masstwof}
\frac{M_\pi}{g} & = & e^{2\gamma/3} \frac{2^{5/6}}{\pi^{1/6}}
\left(\frac{m}{g}\right)^{2/3}\nonumber\\
\left(\frac{M_\eta}{g}\right)^2 & = & \frac{2}{\pi} + 
\left(\frac{M_\pi}{g}\right)^2
\ea 
where $\gamma=0.577216\dots$ is Euler's constant.

\section{Numerical results with the full operator}
\begin{figure}[ht]
\includegraphics[height=6.5cm,width=7.5cm]{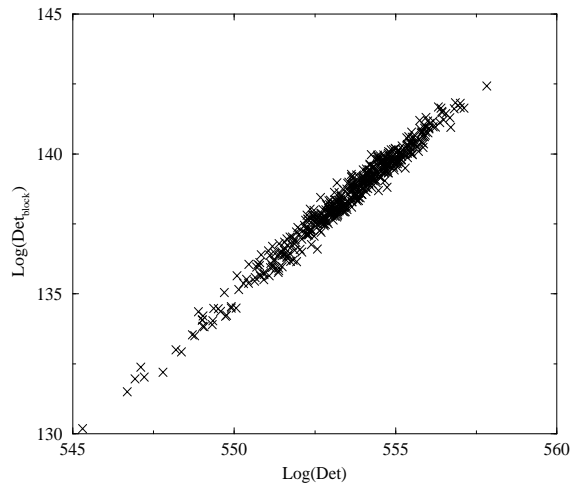}
\vspace{-40pt}
\caption{\label{fig:det}
${\rm Det}(D)$ on the blocked lattice 
versus ${\rm Det} (D)$ on the full lattice.}
\end{figure}

In order to test our approximation method and, at the same time,
to increase the body of informations on the lattice Schwinger model,
we performed an extensive simulation with the exact 
Neuberger-Dirac operator.  We considered the 
Schwinger model at $\beta=6$ on a $24 \times 24$ lattice.
On a lattice of this size, the discretized Dirac operator is
a complex matrix of dimension $1152 \times 1152$, for which
we could use full matrix algebra subroutines without excessive
burden on the resources available to us (Boston
University SGI/Cray Origin 2000 supercomputer). The masses
have been selected on the basis of previous
results which indicated that, for most of the values we
were planning to consider, the lattice would span at least
a few correlation lengths.  We generated 500 independent 
configurations of the gauge variables $U_{\mu}(x)$ distributed 
according to the pure gauge Wilson action of Eq.~(\ref{eq:sg}).
For calculations with one and two
flavors of dynamical fermions, we incorporated the determinant
of the lattice Dirac operator in the averages of the correlation 
functions \cite{unq}.
While with large variations of the determinant this way of proceeding
would lead to an unacceptable variance, in the present
calculation we found the range of values taken by ${\rm Det}(D)$
to be sufficiently limited to warrant our averaging procedure
(see Fig.~\ref{fig:det}).
\begin{table}[htb]
\nonumber
\caption{
Mesons and quark masses for $N_f=2$.}
\begin{tabular}{||l|lll||}
\hline\hline
\multicolumn{4}{||c||}{$\beta=6.0$, $V=24^2$, $N_f=2$} \\
\hline
$m/g$  & $m^{AWI}$ & $m_\pi/g$ & $m_{\eta}/g$ \\
\hline\hline
\multicolumn{4}{||c||}{Analytic} \\
\hline
0 & 0 & 0 & 0.7979  \\ 
\hline
\hline
\multicolumn{4}{||c||}{Full Operator} \\
\hline
      0.00   & -0.004(2)& -0.001(65) & 1.00(25) \\ 
\hline
      0.0244 & 0.0233(2) & 0.14(3) & 1.0(5) \\ 
      0.0485 & 0.0468(6) & 0.26(5) & 1.0(4) \\ 
      0.0960 & 0.0945(13)& 0.44(4) & 1.1(2) \\ 
      0.1427 & 0.1420(14)& 0.57(3) & 1.1(2) \\ 
      0.1884 & 0.1890(13)& 0.69(2) & 1.2(1) \\ 
      0.2333 & 0.2353(11)& 0.79(1) & 1.23(9) \\ 
\hline\hline

\multicolumn{4}{||c||}{Fourier Approximation} \\
\hline
      0.00   & -0.002(2) & -0.005(66)  & -  \\ 
\hline
      0.0244 & 0.018(1) & 0.11(3)      & -  \\ 
      0.0485 & 0.044(1) & 0.24(5)      & -  \\ 
      0.0960 & 0.092(1) & 0.43(4)      & -  \\ 
      0.1427 & 0.139(1) & 0.57(3)      & -  \\ 
      0.1884 & 0.184(1) & 0.68(2)      & -  \\ 
      0.2333 & 0.228(1) & 0.79(1)      & -  \\ 
\hline\hline
\multicolumn{4}{||c||}{Multi-grid Determinant} \\
\hline
      0.00 &  -0.004(1) & -0.002(40)  & 1.09(25) \\ 
\hline
      0.0244 & 0.0234(1) & 0.14(2) & 1.1(4) \\ 
      0.0485 & 0.0469(4) & 0.26(3) & 1.1(4) \\ 
      0.0960 & 0.0946(9) & 0.44(2) & 1.1(2) \\ 
      0.1427 & 0.1420(10)& 0.57(2) & 1.2(2) \\ 
      0.1884 & 0.1889(9) & 0.68(1) & 1.2(1) \\ 
      0.2333 & 0.2351(8) & 0.79(1) & 1.26(9) \\ 
\hline\hline
\end{tabular}
\end{table}

For each configuration, we performed a singular value decomposition
\be
D_W-\frac{1}{a}=U \Lambda \tilde U
\label{eq:svd}
\ee
where $U, \tilde U$ are unitary matrices and $\Lambda$ is diagonal,
real and non-negative.  The operator $V=X/\sqrt{X^{\dagger}X}$ 
in Eq.~\ref{eq:opneub} is then given by
\be
V =U \tilde U
\label{eq:opv}
\ee
We proceeded then to the diagonalization of $V$, calculating
all its eigenvalues and eigenvectors.  From these, it is 
straightforward to calculate both the determinant of the Neuberger-Dirac
operator $D$ as well as its associated propagator $D^{-1}$ for any value 
of the fermion mass $m$.  The full diagonalization of $V$ is computationally
more demanding than the direct calculation of $D^{-1}$, which typically
gives also ${\rm Det} (D)$ as a by-product, but we were interested
in the actual spectrum of $V$ for comparison with the approximations
that will be discussed later.  From the fermion propagators we
calculated the meson propagators projected over zero momentum.
We focused on the two point correlation functions of 
the singlet vector current and the triplet current 
\be
C_{T}(t) = \sum_{x,x',y} \langle V_2(x,y)V_2(x',y+t) \rangle
\label{eq:mcorr}
\ee
because they are saturated by a single particle contribution in
the massless limit. 
We  added to the 
averages the correlators obtained from the interchange of $x$
and $y$ in Eq.~\ref{eq:mcorr} for a further gain in statistics.
\begin{figure}[ht]
\includegraphics[height=7.5cm,width=7.5cm]{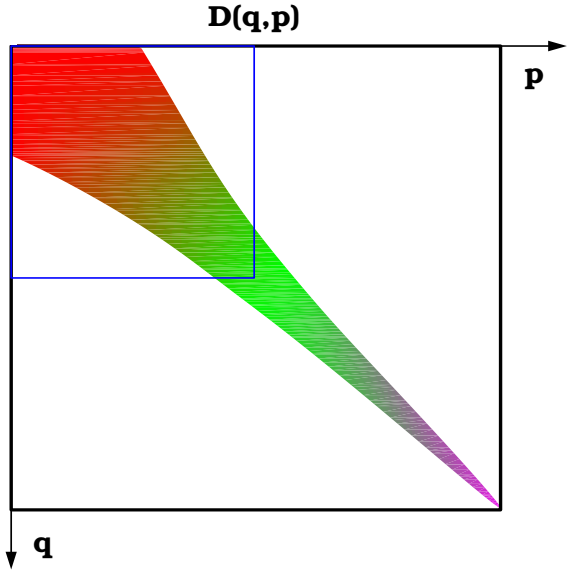}
\vspace{-40pt}
\caption{\label{fig:2rho}
Structure of the Wilson operator in momentum space and 
in a smooth gauge.}
\end{figure} 

From fits to the meson correlators we extracted the meson
masses in a standard manner.
Our results for the meson and quark masses  
as functions of the fermion mass parameter of the action
are reported in Figs.~\ref{fig:mpiNf2} and \ref{fig:rhoNf2}.  
The values we obtained 
for the masses are also reported in the Table.  
The errors have been evaluated
with the jackknife method.  Fig.~\ref{fig:mpiNf2}
illustrates the behavior of the isotriplet mass $M_\pi$ in the model
with two flavors.  This mass is expected to vanish for $m=0$
and chiral perturbation theory predicts an $m^{2/3}$ dependence
on $m$ (see Eq.~\ref{eq:masstwof}). To test this functional
dependence we performed two different fits. First
the pion masses were fitted as 
\be
\frac{M_\pi}{g} = A + B \left(\frac{m}{g}\right)^{2/3}
\ee
using only the highest four quark masses for which $24\times M_\pi>4$.
The results of the fit  $A=-0.001(65)$ and $B=2.10(14)$
confirm that Neuberger's fermions preserve chiral symmetry 
and give values of $B$ in very good agreement with 
the theoretical prediction. The dashed line in Fig.~\ref{fig:mpiNf2}
is the result of the fit. We then 
fitted the pion mass as 
\be
\frac{M_\pi}{g} = B \left(\frac{m}{g}\right)^{\delta}
\ee
obtaining $B=2.10(17)$ and $\delta=0.67(6)$ in perfect agreement 
with the  expected functional dependence. Equally gratifying
is the comparison of the value of the fermion mass $m$ in the Lagrangian
with the value $m^{AWI}$ that can be extracted from the axial Ward
identity Eq.~\ref{eq:AWI}.

In the Table we also report results for the singlet
mass $M_\eta$ versus $m^{4/3}$, always for $N_f=2$.  The singlet meson 
propagator is given by the difference of the connected and disconnected 
terms in the correlator and, because of the cancellations, the errors 
are larger than in the triplet case.  The numerical results are, however, 
consistent with the theoretical prediction of Eq.~\ref{eq:masstwof}.

\section{Approximation in the Fourier Space}
The long range physical properties of the Schwinger Model should be determined 
by the eigenstates of $D$ with eigenvalues in the neighborhood 
of $\lambda=m$ (i.e.~the eigenstates 
of $V$ with eigenvalue $\lambda_V$ closest to $-1$), 
since these are the states with are expected
to go over the physical states of
the continuum in the limit $a \to 0$.  This suggests
that it should be possible to reconstruct 
the physical observables from these states alone.
\begin{figure}[ht]
\includegraphics[height=6.5cm,width=7.5cm]{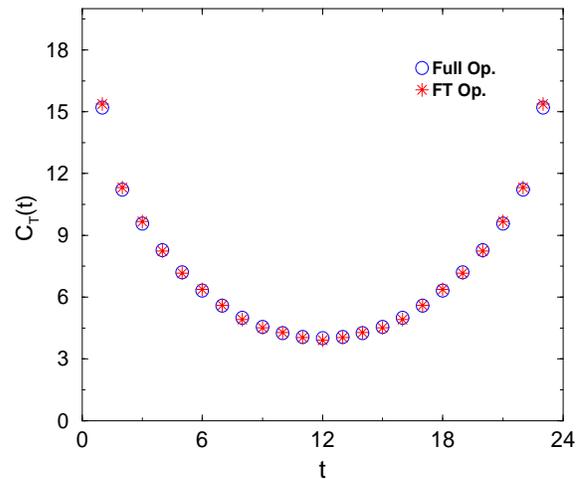}
\vspace{-40pt}
\caption{\label{fig:schw_6.0_1006_TRIP2}
Comparison of the $V_2$-triplet correlation function computed with 
the full operator and the Fourier approximation on a configuration with
$Q=0$.}
\end{figure} 

\begin{figure}[ht]
\begin{center}
\includegraphics[height=6.5cm,width=7.5cm]{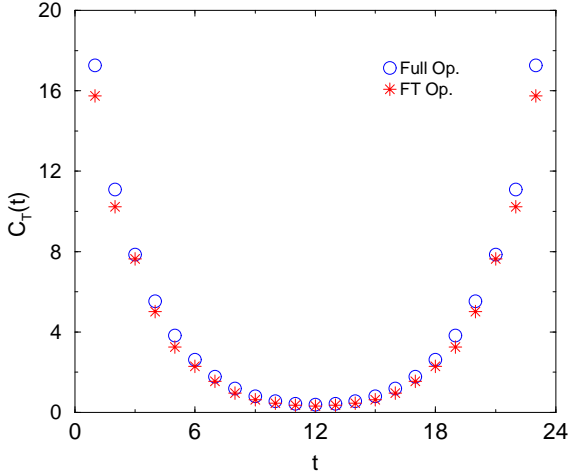}
\vspace{-40pt}
\caption{\label{fig:schw_6.0_1019_TRIP2}
Comparison of the $V_2$-triplet correlation function computed with 
the full operator and the Fourier approximation on a configuration with
$Q=4$.}
\end{center}
\end{figure} 

Here we would like to make the point that the overlap formulation
is particularly well suited for the implementation
of the above approximation, since the unitarity of $V$
provides, in some sense, a constraint on the form of the 
approximation itself.
Of course, one must be careful in attempting any approximation
based on neglecting the short-wavelength part of the spectrum,
even if the corresponding states are largely lattice artifacts,
since one knows that in quantum field theory the infrared
and ultraviolet components of the spectrum are subtly related.
Thus, the chiral eigenstates with $\lambda_V=-1$ (``zero modes''),
which $V$ exhibits in presence of a gauge field with non-trivial topology, 
find their counterpart in states with opposite chirality and $\lambda_V=1$.
\begin{figure}[ht]
\includegraphics[height=6.5cm,width=7.5cm]{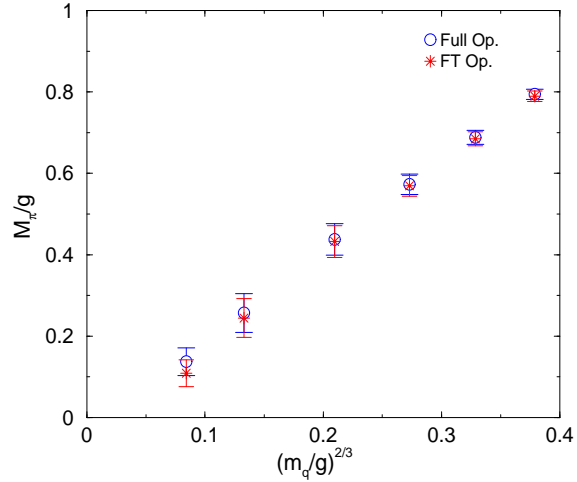}
\vspace{-40pt}
\caption{\label{fig:mpiNf2gfix}
$M_\pi/g$ vs $(m/g)^{2/3}$ for the full operator (circles,blue) 
and the Fourier approximation (stars,red) for $N_f=2$.}
\end{figure} 

\begin{figure}[ht]
\begin{center}
\includegraphics[height=6.0cm,width=6.0cm]{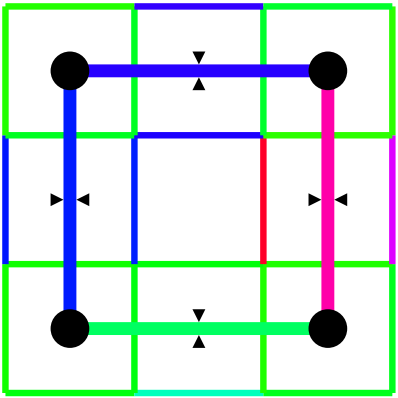}
\vspace{-10pt}
\caption{\label{fig:block}
Sketch of the blocking procedure.}
\end{center}
\end{figure} 
Here too, however, the special features of the overlap formulation
come to the rescue, since, as we will show, both the presence
of zero modes and the correspondence between $\lambda_V=-1$
and $\lambda_V=1$ eigenstates of opposite chirality is
preserved, as we found in our study, by the projection over a subset
of physical states.  

One possible scheme of approximation which is computationally
very convenient consists of performing a projection over states
of low momentum in Fourier space, after gauge fixing to a
smooth gauge field configuration\footnote{A similar but gauge 
invariant procedure is described in \cite{NoiSch} which, however,
turns out to be numerically more expensive.}.  In a smooth gauge, because
of the suppression of short-wavelength fluctuations due to
asymptotic freedom, one expects the structure of the Wilson operator 
to become more and more diagonal in momentum space for increasing
momenta.  This notion is schematically illustrated in Fig.~\ref{fig:2rho}
and also underlies the technique of Fourier acceleration for
the calculation of quark propagators \cite{Wilson}.  
Accordingly, we implemented the following approximation.
For each configuration we factored out the topology, if any, 
as described in \cite{NoiSch} and fixed the Landau 
gauge on the resulting configuration by demanding 
that the function
\be
G = \sum_{x, \mu} {\rm Re}[ U_\mu(x)+U_\mu(x-a\hat\mu)]  
\label{eq:lgauge}
\ee
be maximal.  A relaxation procedure produces several local maxima
(Gribov copies).  Among all these configurations, we selected the
one that produced the maximum for $G$, subject to the
further constraint that for all $x$ and $\mu$ ${\rm Re} U_\mu(x) \ge 0.5$.
Eventually the gauge transformation obtained was applied to the
original configuration. 

Once the gauge has been fixed, the vector space on which the 
fermion matrix $D$ acts is divided into two parts: a long range (LR)
sector spanned by the vectors of the Fourier basis $|p_i\rangle$ ($i=1,\dots n$) with
$p^2_i<p^2_\Lambda$ and its complementary short range (SR) subspace. 
In the LR sector the matrix elements of the Wilson operator are computed
and Neuberger's operator is obtained with the unitarity projection 
in Eq.~(\ref{eq:opneub})
and inverted. In the SR subspace the propagator is approximated
by its free form. The approximated operator is  
$\gamma_5$-Hermitian and it satisfies the GW relation. 
\begin{figure}[ht]
\includegraphics[height=6.5cm,width=7.5cm]{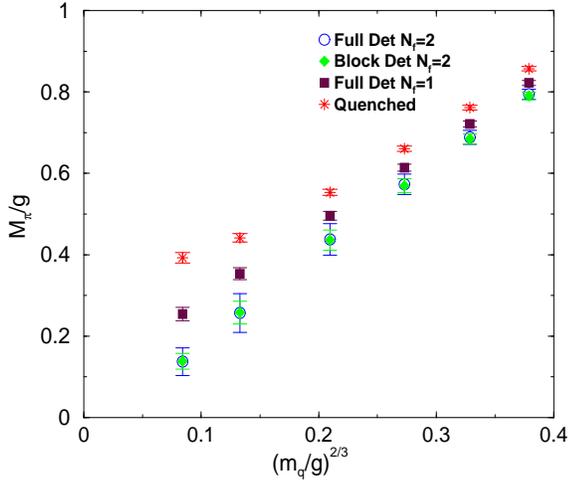}
\vspace{-40pt}
\caption{\label{fig:mpiNf2_grid}
$M_\pi/g$ vs $(m/g)^{2/3}$ quenched (stars,red),
$N_f=1$ (squares,violet) and  $N_f=2$ (circles,blue)
with the full determinant and  $N_f=2$ (diamonds,green)
with the blocked determinant.}
\end{figure} 

In Figs.~\ref{fig:schw_6.0_1006_TRIP2} and (\ref{fig:schw_6.0_1019_TRIP2})
we compare the triplet meson correlation functions $C_T(t)$ obtained with 
the full Neuberger operator and with the approximated one 
for a configuration with trivial topology and for one with the topological charge $Q=4$.
In Fig.~(\ref{fig:mpiNf2gfix}) the triplet meson masses for the full solution and the 
Fourier approximation are reported. The LR subspace used is 1/4 of 
the full space. It is interesting to note that the approximated
solution can be improved if used as preconditioning of the standard algorithms. 

\section{Multi-Grid Approximation for the Determinant}
In the spirit of the long range approximation, another
possibility that can be explored is the 
coarse graining of the gauge field \cite{claudio}.
Starting from an arbitrary gauge configuration on an $N\times N$ lattice, we
apply the blocking procedure sketched in Fig.~\ref{fig:block} 
to obtain a gauge configuration on an $\frac{N}{2}\times \frac{N}{2}$
lattice, which carries all the 
relevant long range features of the large lattice.

The original lattice is divided
into $\frac{N}{2}\times \frac{N}{2}$ blocks of $2^2$ sites.
Each block, after the local Landau gauge is imposed,
corresponds to a site of the coarse lattice. The 
remaining links are averaged as in Fig.~\ref{fig:block} to
give the corresponding links of the coarse lattice.

The absence of additive mass renormalization simplifies the definition of the 
Neuberger operator on the coarse lattice because no fine tuning of the mass
parameter is needed. We calculated the fermion
determinant on the blocked lattice and compared it to the one of the full
lattice as shown in Fig. \ref{fig:det}. 

We incorporated the blocked determinant 
in the averages of the correlation functions.
We compare these results with those 
obtained with the determinant of the full operator. For the pion 
masses the agreement is remarkable
(see Fig.~\ref{fig:mpiNf2_grid}). We did the same for the massive Schwinger 
boson and the results are reported in the Table. Also in this
case, even if the observables are more noisy, the agreement between the
results is remarkable.

The application of these ideas to speed up QCD simulations with  
Neuberger fermions is in progress.


\begin{thebibliography}{9}
\bibitem{HN}
H.~Neuberger,
Phys. Lett. B417 (1998)~141;\\
H.~Neuberger,
Phys. Lett. B427 (1998)~353.
\bibitem{NoiSch}
L. Giusti, C. Hoelbling, C. Rebbi,
in preparation.
\bibitem{GW}
P.~H.~Ginsparg, K.~G.~Wilson, 
Phys. Rev. D25 (1982) 2649.
\bibitem{luscher}
M.~L\"uscher,
Phys. Lett. B428 (1998) 342.
\bibitem{ghl}
C. Gattringer, I. Hip, C. B. Lang,
Phys. Lett. B466 (1999)~287.  
\bibitem{unq}
C.~B.~Lang, T.~K.~Pany, Nucl. Phys. B513 (1998) 645;\\
F.~Farchioni et al, Nucl. Phys. B549 (1999) 364.
\bibitem{Wilson}
G.~Katz et al.,
Phys. Rev. D37 (1988)~1589.
\bibitem{claudio}
R. Brower, R. Edwards, C. Rebbi, E. Vicari,
Nucl. Phys. B336 (1991) 689.
\end{thebibliography}
\end{document}